\documentclass[12pt]{article}
\usepackage{keystroke}
\usepackage{listings}
\usepackage{parcolumns}
\usepackage[sorting=none,backend=bibtex]{biblatex}
\usepackage{tikz}
\usepackage{tikz-qtree}
\usetikzlibrary{trees,calc,shapes.multipart,chains,arrows}

\usepackage{amssymb}
\usepackage{amsmath}
\usepackage{algorithm}
\usepackage{algpseudocode}
\algdef{SE}[DOWHILE]{Do}{doWhile}{\algorithmicdo}[1]{\algorithmicwhile\ #1}%

\usepackage{setspace}

\usepackage{graphicx}

\title{EMFS: Repurposing SMTP and IMAP for Data Storage and Synchronization}
\author{
	William Woodruff \\
	\href{mailto:william@tuffbizz.com}{william@tuffbizz.com}
}
\date{
	\today \\
	v1.0
}

\begin{document}
\maketitle

\begin{abstract}
\noindent Cloud storage has become a massive and lucrative business, with
companies like Apple, Microsoft, Google, and Dropbox providing hundreds of
millions of clients with synchronized and redundant storage. These services
often command price-to-storage ratios significantly higher than the market
rate for physical storage, as well as increase the surface area for data
leakage. In place of this consumer-unfriendly status quo, I propose using widely
available, well standardized email protocols like SMTP and IMAP in
conjunction with free email service providers to store, synchronize, and share
files across discrete systems.
\end{abstract}

\newpage

\begin{spacing}{0.95}
\tableofcontents
\end{spacing}

\newpage

\section{Introduction}
Remote storage and synchronization of data, commonly referred to as ``cloud"
storage, has become increasingly popular with companies and consumers alike
as both a redundancy measure and as a way to share informations across a diverse
range of platforms. For companies, this reduces or eliminates the need to
maintain internal network filesystems like NFS and CIFS. For consumers, having
a remotely accessible copy of data simplifies common tasks like social sharing
(pictures, video) and reduces the frustration of using multiple computers
(desktop, laptop, smartphone) for a common task.

Cloud storage providers (CSPs) generally divide their service into a free
tier and a paid tier. Free tier users are usually restricted in terms of how
much they can upload, but may also be subject to slower upload speeds and
limited trial periods. Paid users are also usually restricted in terms of total
upload capacity, but at much higher limits and may not be subject to the same
level of throttling as free users.

Beyond price concerns, the prevalence of CSPs has expanded the surface area for
data breaches on both an individual and company level. Incidents like the
Apple-owned iCloud leak \cite{Apple2014} and security loopholes like Dropbox's
lax public URL policy \cite{Cluley2014} reflect poor security choices by users
and service providers alike. This is exacerbated by the widespread use of
proprietary (and mutually incompatible) clients and protocols by CSPs, leaving
users at the mercy of their provider of choice for updates.

There have been attempts to provide open-source and user-controlled alternatives
to CSPs \cite{Syncthing, ownCloud}, but all rely on a level of technical
commitment and expertise comparable to managing a traditional networked
filesystem. Since the goal of ``cloud" storage is to eliminate the technical
barrier to data synchronization, these alternatives are not currently suitable
for the majority of CSP users.

As an alternative to both traditional CSPs and their open-source replacements,
I propose a system that uses the already open Simple Mail Transport Protocol
(SMTP) \cite{rfc821} and Internet Message Access Protocol (IMAP) \cite{rfc3501}
in conjunction with free email service providers (ESPs) to store, synchronize,
and share data between both users and machines. I call this system the
\textit{\underline{EM}ail \underline{F}ile\underline{S}ystem}, or \textit{EMFS}.

\section{Email as a General-Purpose Storage Service}

\subsection{Usenet as a Precedent}

\subsubsection{Protocol and Topological Similarities}
The topology of the Usenet network bears an uncanny resemblance to the
contemporary email network, with clients connecting to a ring of relay servers
responsible for distributing the latest posts to all recipient newsgroups.

The Network News Transport Protocol (NNTP) \cite{rfc977, rfc3977}, the dominant
contemporary Usenet protocol, also closely resembles SMTP in
message structure:

\noindent\begin{minipage}{.49\textwidth}
\begin{lstlisting}[caption=NNTP session]
< CONNECT news.foo.com
> 200 news.foo.com NEWS
< POST
> 340 Ok, recommended ID <beef@news.foo.com>
< From: bar@baz.com
< Newsgroups: misc.quux
< Subject: Hello!
< Message-ID: <beef@news.foo.com>
<(*\Return*)
< Hello, World!
< .
> 240 article posted ok
< QUIT
> 205 Goodbye
\end{lstlisting}
\end{minipage}\hfill
\begin{minipage}{.49\textwidth}
\begin{lstlisting}[caption=SMTP session]
< CONNECT smtp.foo.com
> 220 smtp.foo.com SMTP
< HELO smtp.foo.com
> 250 Hello!
< MAIL FROM: <bar@baz.com>
> 250 Ok
< RCPT TO: <quux@baz.com>
> 250 Ok
< DATA
> 354 End data with <CR><LF>.<CR><LF>
< Subject: Hello!
< From: bar@baz.com
< To: quux@baz.com
<(*\Return*)
< Hello, World!
< .
> 250 Ok: queued as 105
< QUIT
> 221 Goodbye
\end{lstlisting}
\end{minipage}\hfill

\subsubsection{Usage Similarities}

Usenet was originally developed to share text-based news and threaded
discussions. As academic and social usage of the Internet expanded,
Usenet users took advantage of the network's mirroring capabilities to share
encoded binary files. The most notable Usenet hierarchy for binary sharing
has historically been \texttt{alt.binaries.*}, with many providers and
authorities limiting access to their entire \texttt{alt.*} hierarchy due to the
abundance of illegal content \cite{cnet2008:01, cnet2008:02}. As of 2010,
99\% of all traffic over NNTP is yEnc binary data \cite{gi2010}.
Because Usenet mirrors its data across every provider it has
also been used as a backup service \cite{ubackup}, with users uploading
encrypted data to globally visible newsgroups.

Email, like Usenet, was originally developed to share text-based messages,
albeit between sets of addresses instead of discussion hierarchies.
Just like Usenet, email adapted to share binary content as the needs of users
diversified to include multimedia and rich document types. Unlike Usenet, email
networks have never seen the proliferation of organized binary sharing.

\subsection{Advantages over NNTP}

There are a number of benefits to storing and sharing binary data over SMTP
and IMAP instead of NNTP.

Unlike Usenet messages, email messages are not directed to a global hierarchy.
The destination of an individual (SMTP composed) email may be a mailing list,
$N$ distinct addresses, or even the sending address itself. As a result, clients
may exercise more autonomy over where their data is sent and when it requires
encryption.

Although yEnc has emerged as a de-facto standard \cite{gi2010} encoding on
Usenet, there is no \textit{formal} standard for sending non-text data over
NNTP. Consequently, Usenet newsreaders regularly suffer both false negatives
and positives when fetching encoded binaries \cite{claus2002, nixon2002}. By
comparison, MIME-encoded binaries were introduced as an extension in 2000 and
are now universally handled by rich email clients \cite{rfc3030}.

\subsection{Advantages over CSPs}

\subsubsection{Security}
\label{security}

The use of an extant email account for storage confers a number of security and
usability benefits compared to a traditional CSP. Instead of having to register
for a new service, users can access their files with their email credentials.
This alone eliminates the attack surface normally associated with
cloud storage, with no danger of data leakage from public URLs and half as many
potential points of misconfiguration or attack.

The prevalence of TLS-secured open email protocols is also advantageous,
as \textit{EMFS} may take advantage of the existing email security and privacy
infrastructure instead of implementing its own. In particular, STARTTLS for both
SMTP and IMAP \cite{rfc2487, rfc2595} is universally supported by large email
providers. Beyond transit security, an implementation of the open Pretty
Good Privacy (PGP) standard \cite{rfc4880} can be used to ensure storage
security.

\subsubsection{Extensibility and Transparency}

To protect their intellectual property and remain competitive, large CSPs make
extensive use of proprietary protocols and clients. Although these methodologies
can be observed and analyzed \cite{Drago:2012:IDU:2398776.2398827}, their closed
nature hinders interoperability and locks users into vendor-specific ecosystems.
Because \textit{EMFS} operates solely on the email infrastructure and utilizes
open protocols, it can be extended and modified trivially both by users and
developers as demands evolve.

The presence of \textit{EMFS} on a user's account is also fully opaque to
the ESP, as \textit{EMFS} traffic is identical to normal email traffic.
From the user's perspective, the only evidence of \textit{EMFS} is an IMAP
folder of their choosing used to store messages. In the full spirit of the SMTP
and IMAP protocols, this allows the user to treat their email client as a basic
file manager by using standard email primitives (Compose, Edit, Delete) to
manipulate synchronized data.

\section{Architecture}

Like any other filesystem, the architecture of \textit{EMFS} can be broken into
discrete primitives that can be categorized by type or operation.
\textit{EMFS} is also interacted with as a normal hierarchical filesystem,
providing a tree whose root is the virtual mountpoint for the \textit{EMFS}
instance.

\subsection{Filesystem Primitives}

\subsubsection{Types}

\textit{EMFS} is aware of two primitive types: files and directories.

\paragraph{Files}

\textit{EMFS} files are chains of $N$ email messages, both header and body,
where $N$ is greater than or equal to $1$. In terms of lookup, a message chain
for a given file $F$ divided into $N$ messages behaves like a linked list:

\begin{tikzpicture}[list/.style={rectangle split, rectangle split parts=2, draw, rectangle split horizontal}, >=stealth, start chain]
	\node[list,on chain] (A) {$F_{0}$};
	\node[list,on chain] (B) {$F_{1}$};
	\node[list,on chain] (C) {$\cdots$};
	\node[list,on chain] (D) {$F_{N - 2}$};
	\node[list,on chain] (E) {$F_{N - 1}$};
	\node[on chain,draw,inner sep=6pt] (F) {};
	\draw (F.north east) -- (F.south west);
	\draw (F.north west) -- (F.south east);
	\draw[*->] let \p1 = (A.two), \p2 = (A.center) in (\x1,\y2) -- (B);
	\draw[*->] let \p1 = (B.two), \p2 = (B.center) in (\x1,\y2) -- (C);
	\draw[*->] let \p1 = (C.two), \p2 = (C.center) in (\x1,\y2) -- (D);
	\draw[*->] let \p1 = (D.two), \p2 = (D.center) in (\x1,\y2) -- (E);
	\draw[*->] let \p1 = (E.two), \p2 = (E.center) in (\x1,\y2) -- (F);
\end{tikzpicture}

Message chains represent literal data, with no facilities for UNIX-like
symbolic linking. Two functions are required to generate the content in an
\textit{EMFS} message chain, an 8-bit encoding function $\mathit{Encode8}$ and a
hashing function $\mathit{Hash}$. $\mathit{Hash}$ may be as simple as an
iterative function.

An \textit{EMFS} file might have a message-representation as follows:

\begin{lstlisting}[caption=EMFS Message,numbers=left,showspaces=true,label={messagerep}]
From: foo@bar.com
To: foo@bar.com
Subject: first-id-hash filename
EMFS-Filename: filename
EMFS-Next: next-id-hash

encoded-body
\end{lstlisting}

Where \texttt{first-id-hash} is the first ID hash in the sequence,
\texttt{next-id-hash} is the ID hash of the next message in $F$, and the
generic \texttt{id-hash} is found by:

$$
\texttt{id-hash} \gets \mathit{Hash}(\texttt{filename}, \mathit{N})
$$

And where \texttt{encoded-body} is found by:

$$
\texttt{encoded-body} \gets \mathit{Encode8}(\texttt{file-slice})
$$

Where \texttt{file-slice} is the array of data of $F_{N}$.

\paragraph{Directories}

\textit{EMFS} directories are mapped directly onto the IMAP notion of
``folders". An IMAP folder is said to contain $M$ messages for $K$ files, each
file split across $K_{N}$ messages such that:

$$
M = \sum_{F=1}^{K} F_{N}
$$

As such, from the IMAP perspective, the \textit{EMFS} hierarchy looks like this:

\begin{tikzpicture}[grow'=right,level distance=1.25in,sibling distance=.25in]
\tikzset{edge from parent/.style=
			{thick, draw, edge from parent fork right},
		 every tree node/.style=
			{draw,minimum width=1in,text width=1in,align=center}}
\Tree
	[.\texttt{emfs/}
		[.{\texttt{example/}}
			[.$\mathit{Hash}(K'_{0_{0}})$\ \texttt{hello.mp4} ]
			[.$\mathit{Hash}(K'_{0_{1}})$\ \texttt{hello.mp4} ]
			[.$\mathit{Hash}(K'_{0_{2}})$\ \texttt{hello.mp4} ]
		]
		[.$\mathit{Hash}(K_{0_{0}})$\ \texttt{example.txt} ]
		[.$\mathit{Hash}(K_{1_{0}})$\ \texttt{example.png} ]
	]
\end{tikzpicture}

And from the file manager's perspective:

\begin{tikzpicture}[grow'=right,level distance=1.25in,sibling distance=.25in]
\tikzset{edge from parent/.style=
			{thick, draw, edge from parent fork right},
		 every tree node/.style=
			{draw,minimum width=1in,text width=1in,align=center}}
\Tree
	[.\texttt{emfs/}
		[.{\texttt{example/}}
			[.\texttt{hello.mp4} ]
		]
		[.\texttt{example.txt} ]
		[.\texttt{example.png} ]
	]
\end{tikzpicture}

Files with only one message (i.e., $F_{N} \mid N = 1$) are constituted directly.
More notably, \texttt{hello.mp4} is constituted into a single file from
three messages (i.e., $F_{N} \mid N = 3$) under the IMAP \texttt{example/} folder.

\subsubsection{Operations}

\paragraph{Filesystem Creation}

Because the entirety of each discrete \textit{EMFS} instance resides in its
own IMAP folder hierarchy, the creation of an instance requires the creation of
a root IMAP folder. This is accomplished by issuing a \texttt{CREATE} verb
during a short IMAP session:

\begin{center}
\begin{tikzpicture}[>=latex]
\coordinate (A) at (2,5);
\coordinate (B) at (2,0);
\coordinate (C) at (6,5);
\coordinate (D) at (6,0);
\draw[thick] (A)--(B) (C)--(D);
\draw (A) node[above]{\Large Client};
\draw (C) node[above]{\Large Server};

\coordinate (E) at ($(A)!.1!(B)$);
\draw (E) node[left]{$\mathit{EMFS\_Init}$};

\coordinate (F) at ($(C)!.25!(D)$);
\draw (F) node[right]{$\mathit{IMAP\_Auth}$};
\draw[->] (E) -- (F) node[midway,sloped,above]{\texttt{LOGIN U P}};

\coordinate (G) at ($(A)!.4!(B)$);
\draw (G) node[left]{$\mathit{EMFS\_Create\_Root}$};
\draw[->] (F) -- (G) node[midway,sloped,above]{\texttt{OK LOGIN}};

\coordinate (H) at ($(C)!.6!(D)$);
\draw (H) node[right]{$\mathit{IMAP\_Create}$};
\draw[->] (G) -- (H) node[midway,sloped,above]{\texttt{CREATE EMFS}};

\coordinate (I) at ($(A)!.75!(B)$);
\draw (I) node[left]{$\mathit{EMFS\_Logout}$};
\draw[->] (H) -- (I) node[midway,sloped,above]{\texttt{OK}};

\coordinate (J) at ($(C)!.9!(D)$);
\draw (J) node[right]{$\mathit{IMAP\_Logout}$};
\draw[->] (I) -- (J) node[midway,sloped,above]{\verb$LOGOUT$};
\end{tikzpicture}
\end{center}

\paragraph{Directory Creation}

Like the creation of the \textit{EMFS} instance, the creation of individual
directories within an instance relies on the IMAP \texttt{CREATE} verb. To
create nested directories below the root level in a fashion similar to UNIX's
\texttt{mkdir -p}, the \texttt{SELECT} verb is also required. Shown iteratively:

\begin{algorithm}
\caption{Directory Creation}
\begin{algorithmic}[1]
\Procedure{EMFS\_Mkdir}{$\mathtt{directory}$}
\State $\mathtt{delimiter} \gets \text{the symbol used to delimit directories}$
\State $\mathtt{directories} \gets \mathit{Split}(\mathtt{directory}, \mathtt{delimiter})$
\For{$\mathtt{d} \in \mathtt{directories}$}
\If{$\nexists\mathtt{d}$}
\State $\mathit{IMAP\_Create}(\mathtt{d})$
\EndIf
\State $\mathit{IMAP\_Select}(\mathtt{d})$
\EndFor
\EndProcedure
\end{algorithmic}
\end{algorithm}

\paragraph{Directory Deletion}

In the most extreme case, an \textit{EMFS} directory deletion should remove
all messages and subfolders in the corresponding IMAP folder in a fashion
similar to UNIX's \texttt{rm -rf}. Because the IMAP \texttt{DELETE} verb
will refuse to operate when given a folder with subfolders, \texttt{EMFS} must
descend to all subfolders and perform \texttt{DELETE} on them first. Shown
recursively:

\begin{algorithm}
\caption{Directory Deletion}
\begin{algorithmic}[1]
\Procedure{EMFS\_Rmdir}{$\mathtt{directory}$}
\State $\mathtt{subdirs} \gets \text{all subdirectories of } \mathtt{directory}$
\If{$\mathtt{subdir} = \varnothing$}
\State $\mathit{IMAP\_Delete}(\texttt{directory})$
\State \textbf{return}
\Else
\For{$\mathtt{d} \in \mathtt{subdirs}$}
\State $\mathit{EMFS\_Rmdir}(\textit{d})$
\EndFor
\EndIf
\EndProcedure
\end{algorithmic}
\end{algorithm}

\paragraph{File Creation}

The conversion of file data into a chain of messages is complicated by the lack
of a standardized encoded message size limit across common mail servers
\cite{google:01, yahoo:01, microsoft:01}. For the sake of generality, the
\textit{EMFS} message chunking algorithm makes reference to this size limit as
$S$.

\begin{algorithm}
\caption{File Creation}
\begin{algorithmic}[1]
\Procedure{EMFS\_Put}{$\mathtt{filename}$}
\State $\mathtt{file} \gets \mathit{Encode8}(\mathit{Read}(\mathtt(filename)))$
\State $\mathtt{slices} \gets$ file slices of size $\leq \mathit{S}$ of $\mathtt{file}$
\State $\mathtt{messages} \gets \mathit{EMFS\_Pack}(\mathtt{filename}, \mathtt{slices})$
\For{$\mathtt{m} \in \mathtt{messages}$}
\State $\mathit{SMTP\_Send}(\mathtt{m})$
\EndFor
\EndProcedure
\end{algorithmic}
\end{algorithm}

Where $\mathit{EMFS\_Pack}$ is defined as follows:

\begin{algorithm}
\caption{Message Packing}
\begin{algorithmic}[1]
\Procedure{EMFS\_Pack}{$\mathtt{filename}$, $\mathtt{slices}$}
\State $\mathtt{messages} \gets$ an empty list
\State $\mathtt{size} \gets \mathit{Count}(\mathtt{slices}) - 1$
\For{$\mathtt{i} \in \mathit{Range}(0, \mathtt{size})$}
\State $\mathtt{id} \gets \mathit{Hash}(\mathtt{slices}[\mathtt{i}])$
\State $\texttt{next\_id} \gets -1$
\If{$\mathtt{i} < \mathtt{size}$}
\State $\mathtt{next\_id} \gets \mathit{Hash}(\mathtt{slices}[\mathtt{i} + 1])$
\EndIf
\State $\mathtt{message} \gets \mathit{Build\_Message}(\mathtt{filename}, \mathtt{id}, \mathtt{next\_id}, \mathtt{slices}[\mathtt{i}])$
\State $\mathit{Append}(\mathtt{messages}, \mathtt{message})$
\EndFor
\State \textbf{return} $\mathtt{messages}$
\EndProcedure
\end{algorithmic}
\end{algorithm}

Where $\mathit{Build\_Message}$ constructs an SMTP envelope and body of the form
specified in Listing \ref{messagerep}.

\paragraph{File Deletion}

File deletion is a straightforward matter of following the hash chain after
resolving the first node from the \texttt{EMFS-Filename} SMTP envelope header
field.

\begin{algorithm}
\caption{File Deletion}
\begin{algorithmic}[1]
\Procedure{EMFS\_Delete}{$\mathtt{filename}$}
\State $\mathtt{message} \gets $ the first SMTP message in the file chain
\Do
\State $\mathtt{next\_id} \gets \mathit{SMTP\_Header}(\mathtt{message}, \texttt{EMFS-Next})$
\State $\mathit{IMAP\_Delete}(\mathtt{message})$
\State $\mathtt{message} \gets \mathit{EMFS\_Next}(\mathtt{filename}, \mathtt{next\_id})$
\doWhile{$\mathtt{next\_id} \neq -1$}
\EndProcedure
\end{algorithmic}
\end{algorithm}

\paragraph{Indexing}

Because operations that require file or directory discovery would become
extremely expensive in terms of both network time and computation if each
performed its own IMAP \texttt{LIST} verb, \textit{EMFS} maintains its own
cached index of the IMAP message hierarchy. This index is built at the beginning
of each session and updated whenever an \textit{EMFS} operation modifies the
IMAP hierarchy.

\section{Applying \textit{EMFS} to Common ESPs}

\subsection{ESP Statistics}

\subsubsection{Google Gmail}

Gmail is the world's largest free ESP, hosting over 900 million active users
as of May 2015 \cite{gmail:01}. It was also one of the first to offer large
storage capacities for nonpaying users, beginning with 1GB and currently
offering 15GB shared across the all services tied to a user's Google account
\cite{kuchinskas, gmail:02}. Gmail caps attachment size to 25MB
\cite{google:01}.

\subsubsection{Microsoft Outlook}

Microsoft Outlook, previously Hotmail, is one of the earliest web-based
ESPs. It is also currently the second largest ESP with a free plan, with 400
million active as of 2014 \cite{microsoft:02}. Outlook's free plan includes
unlimited email storage and is not tied to quotas for other services on the same
account \cite{microsoft:03}. Outlook caps attachment size to 20MB
\cite{microsoft:01}.

\subsubsection{Shared Characteristics}

Together, Gmail and Outlook store the emails of over 1.3 billion active users.
They are both core components of mature corporations, and both have
established themselves as standards for both personal and corporate email
service \cite{frommer:01}.

In addition to their webmail interfaces, both give their users full SMTP and
IMAPv4 access, including support for session encryption via TLS. Both offer
large attachment sizes and large storage capacities (albeit ultimately limited
in the case of Gmail). Together these qualities make Gmail and Outlook, as
well as other large ESPs like Yahoo!\@ and AOL, more than suitable as hosts
for \textit{EMFS} instances.

\subsection{Potential Hurdles and Concerns}

\subsubsection{Countermeasures by ESPs}

Because \textit{EMFS} takes advantage of the storage offerings of ESPs without
sending meaningful amounts of email traffic between distinct users, it's
possible that large providers will take steps to curb such usage of their
service. In addition, since many large ESPs are also CSPs, usage of
\textit{EMFS} on their free accounts may be treated as an attempt to circumvent
payment for their services.

\subsubsection{Sharing Between Users}

Although IMAP allows email users to create folder hierarchies in their accounts
and organize their messages into these folders, all ESPs employ a ``standard"
behavior of sending new incoming emails to the user's inbox folder. This
behavior is in conflict with \textit{EMFS}'s principles of invisibility and
noninteraction with regards to ``normal" email traffic. Although \textit{EMFS}
can veil this behavior on an individual email account by issuing IMAP commands
to move \textit{EMFS} messages to their dedicated hierarchy, sharing attempts
between multiple email addresses may be complicated by a need to initially
index all received mail for files sent before \textit{EMFS} configuration
by the recipient.

\section{Conclusions}

\textit{EMFS} has many advantages over conventional CSPs, as well as some
disadvantages.

In principle, \textit{EMFS} is substantially simpler to use and secure than
any CSP. Its simplicity derives from its use of the user's email account and
credentials rather than a distinct account and credentials on a CSP, meaning
that the user need only remember one login to access both files and emails.
``Creation" of an \textit{EMFS} store is the simple act of signing in to the
client for the first time. It is also secured on a transport level by ubiquitous
STARTTLS support in both SMTP and IMAP among large ESPs. Overall, the
combination of simplicity in setup and a security ecosystem built on open
standards makes \textit{EMFS} an appealing alternative to distinct CSP accounts
and unaudited encryption stacks.

In addition to its simplicity of operation, \textit{EMFS} is also completely
vendor independent. It makes a minimum number of assumptions about the
capabilities of the servers it uses, allowing users to be flexible in their
ESP choices. Although its structure is inspired by a desire to store
data at no cost to the user, paid email plans are just as capable of hosting it.

With these advantages come some natural disadvantages. \textit{EMFS} makes
no attempt to provide a simple user-to-user synchronization mechanism equivalent
to the ``shared folder" idiom in cloud storage. It also does not provide an
HTTP gateway for one-way file sharing, as many CSPs do. Neither of these tasks
is impossible under the \textit{EMFS} infrastructure, but both incur significant
complexities and consideration in setup. \textit{EMFS} performance is also
heavily dependent on the good grace and bandwidth of the user's ESP.

Overall, \textit{EMFS} is best suited to the personal storage and
synchronization needs of a single user across multiple machines. Portability
and platform support is limited only by the existence of a storage medium and
a functional network stack.

\section{Afternotes}

\subsection{Future Directions}

The potential use of PGP to ensure message security was noted in Section
\ref{security}, but no approach was given. In principle, adding PGP to the
\textit{EMFS} architecture should be as simple as adding an encryption and
decryption layer before and after each individual message transmission and
retrieval.

A compression layer is another potential addition to \textit{EMFS}, either
applied early to the whole data being uploaded or after chunking (but before
encryption). The application of this would be a performance tradeoff between
(de-)compression times and long message chains. Depending on the frequency of
access, performance of IMAP calls, and choice of hardware and compression
algorithm, it may or may not be preferable to limit the length of message
chains via such a layer.

Although this paper lays out algorithms for the most common \textit{EMFS}
operations, it does not provide much detail on aspects of synchronization
from the user's perspective. The simplest potential synchronization mechanism
would be an file event watcher, pointed at the directory of the \textit{EMFS}
``mount". As file events are observed, the appropriate \textit{EMFS} operation
could be dispatched. This has the benefit of being divorced from the
heavy lifting done by \textit{EMFS} in chunking and reconstituting messages
into files.

\subsection{Implementation}

The first program resembling an \textit{EMFS} implementation was written (well)
before this paper, over a period of 48 hours at a hackathon. It lacks many of
the abilities of the system described here, but provides a brief glimpse at a
full user experience in the form of both web and virtual filesystem frontends.
It can be found at \url{https://github.com/fcf634cbe8298176f7c576faed0e500a},
although the author does not advise its usage.

\begin{figure}[h]
\centering
\includegraphics[width=0.9\textwidth]{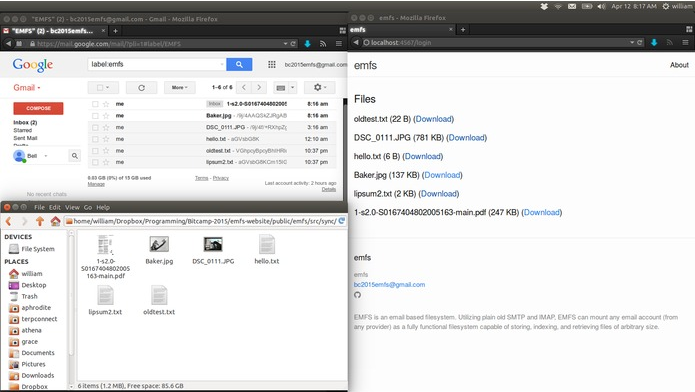}
\caption{The earliest \textit{EMFS}, showing email, filesystem, and web views.}
\end{figure}

A full implementation corresponding more closely to the system described here
is in progress, but has not been released yet. When released, it will be
available at \url{https://github.com/emfs-redux}.

\newpage

\printbibliography

\end{document}